\documentclass[a4paper,10pt]{article}

\usepackage{bm,amsmath,amssymb,graphicx,mathrsfs}

\newcommand{\nc}{\newcommand}
\nc{\rnc}{\renewcommand}
\nc{\nn}{\nonumber}
\nc{\bra}{\langle}
\nc{\ket}{\rangle}
\rnc{\(}{\left(}
\rnc{\)}{\right)}
\rnc{\[}{\left[}
\rnc{\]}{\right]}

\nc{\sh}{\mathrm{sh}}

\begin{document}

\title{On Baxter's $Q$ operator of the higher spin XXZ chain
at the Razumov-Stroganov point}

\author{Kohei Motegi\thanks{E-mail: motegi@gokutan.c.u-tokyo.ac.jp} \\
\it Okayama Institute for Quantum Physics, \\
\it Kyoyama 1-9-1,
Okayama 700-0015, Japan}

\date{\today}

\maketitle

Based on the conjecture for the exact eigenvalue of the transfer matrix
of the higher half-integer spin XXZ chain at the Razumov-Stroganov point,
we evaluate the corresponding Baxter's $Q$ operator in closed form
by solving the $TQ$ equation.
The combination of the $Q$ operators on the ``right side" and the ``wrong side"
is shown to produce the hierarchy of functional relations.

\section{Introduction}
Among the various methods to analyze one-dimensional quantum integrable models,
the Bethe ansatz is one of the traditional and most powerful methods.
There are also many variants of the Bethe ansatz itself today.
The coordinate Bethe ansatz
was invented by Bethe himself to diagonalize the
Hamiltonian of the Heisenberg XXX chain, leading to the derivation
of the Bethe ansatz equation \cite{Bethe}.
Later on, alternative techniques to obtain the
Bethe ansatz equation developed, such as
the algebraic Bethe ansatz (quantum inverse scattering method)
\cite{FSD,KBI} and the analytic Bethe ansatz \cite{BR1,KS}.
The spirit of these new types of the Bethe ansatz
is to diagonalize the transfer matrix associated with the model,
instead of dealing the Hamiltonian directly.

One version of the Bethe ansatz invented by Baxter now called the
$Q$ operator method \cite{Baxter1} is a way to diagonalize the transfer matrix
by making gauge transformation with the help of the $Q$ operator.
The so-called $TQ$ equation is essentially the
Bethe ansatz equation. 
By setting the spectral parameter to the zeros of the $Q$ operator
turns the $TQ$ equation to the Bethe ansatz equation,
which implies the zeros of the $Q$ operator are the 
Bethe roots, i.e.,
the $Q$ operator contains the information of the eigenstates
of the model.
Among the works on the $TQ$ equation and the $Q$ operator
\cite{AF,KLWZ,BLZ1,BLZ2,BLZ3,DT,PS,BS,RW,Korff,BM,YNZ,Roan,
FM,BDKM,Kojima,BT,Tsuboi,BLMS,KLT,BGKNR},
there has been a large advance on the
representation theoretical construction
for the last twenty years.
One of its motivations was to apply this method
originally developed for the integrable spin chains
to conformal field theory.
It also gained interest recently from the particle physics
to check the validity of the AdS/CFT correspondence,
the conjectural duality between a string theory with gravity
on the anit de Sitter apace and a gauge theory on its boundary,
from the gauge theory side.

Progresses in the $TQ$ equation
and the $Q$ operator on the XXZ chain have also been made.
One of them is the speciality at some point of the anisotropy parameter.
Taking the anisotropy parameter to the special point
now called the Razumov-Stroganov point,
the transfer matrix eigenvalue corresponding to the
ground state was conjectured to be expressed in an
particularly simple form.
Using the exact form of the transfer matrix \cite{Baxter2},
the $Q$ operator was expressed in a explicit form \cite{FSZ,St,RS1}.
Furthermore, various exact quantities
on the groundstate components and correlation functions
of the XXZ chain were conjectured to be related with
those of the alternating sign matrix \cite{RS2,BdeGN,deGBNM}.
The symmetries of the partition function
of the six vertex model was also clarified \cite{Okada,St2}, and
there is an extensive study of the related loop models
by making using of the $q$-KZ equations
\cite{DiFrancesco,DZZ,RSZ,CS,Cantini} for example.

In this paper, we generalize some of the analysis previously done for the
spin-1/2 XXZ chain at the Razumov-Stroganov point to higher spins.
Contrary to the spin-1/2 case, not is much known for the higher spins
at the Razumov-Stroganov point.
Several works of them are:
the exact eigenvalue of the transfer matrix of the higher spin XXZ chain 
\cite{DST},
the sum rule of the fused loop model \cite{ZinnJustin},
some groundstate components for the spin-1 XXZ chain \cite{Hagendorf} and
the Macdonald polynomial description of the partition function of the
fused vertex model \cite{FB}.
Starting from the conjectured eigenvalue of the transfer matrix \cite{DST},
we investigate the properties of
the $TQ$ equation and the $Q$ operators of the higher half-integer
spin XXZ chain.
We first analyze the $Q$ operator
by solving the $TQ$ equations in two ways.
First, by expanding the $Q$ operators in terms of the symmetric polynomials
of the Bethe roots, we show that to solve the $TQ$ equation
at the Razumov-Stroganov point is to solve a set of linear equations
with the symmetric polynomials as unknown parameters.
Next, we evaluate the $Q$ operator in closed form
by use of the interpolation formula, generalizing
the case for the spin-1/2 XXZ chain \cite{St}.
Furthermore, the combination of the two $Q$ operators into the form
of discrete Wronskian is shown to produce a hierarchy of functional
relations \cite{KR,BR2,KP,KNS}.

This paper is organized as follows.
In the next section,
we review the $TQ$ equation and its explicit
form of the transfer matrix eigenvalue at the Razumov-Stroganov point,
and make an analysis on the
$Q$ operator first by making the symmetric polynomial expansion
of the $TQ$ equation. We next evaluate the $Q$ operator
in closed form by use of the interpolation formula in section 3,
taking the spin-3/2 chain for example. The results for the generic
half-integer spin chain is presented in section 4.
In section 5, we investigate the functional relations.
Section 6 is devoted to the summary of this paper.

\section{Symmetric Polynomial Expansion}
Our starting point of the analysis is the
Baxter's $TQ$ equation of the integrable higher spin-$s$ XXZ chain
with $M$ sites under the periodic boundary condition
\begin{align}
T(u)Q(u)=\mathrm{sh}^{M}(u+s \eta) Q(u-\eta)
+\mathrm{sh}^{M}(u-s \eta) Q(u+\eta),
\label{TQ}
\end{align}
where  $T(u)$ is the eigenvalue of the transfer matrix $\hat{T}(u)$
whose auxiliary space has spin-$1/2$ and
the quantum space is the $M$-fold tensor product of spin-$s$ spaces.
$u$ is the spectral parameter and
$\eta$ is the anisotropy parameter associated with the XXZ chain
($\eta=0$ corresponds to the isotropic XXX point).
$Q(u)$ is the eigenvalue of the $Q$ operator $\hat{Q}(u)$ \cite{Baxter1}
which was introduced to diagonalize the transfer matrix.
$Q(u)=\prod_{j=1}^p \mathrm{sh}(u-u_j)$ encodes the information of the
eigenstate of the transfer matrix $\hat{T}(u)$.
Indeed, setting the spectral parameter $u$ to $u=u_j$,
the $TQ$ equation \eqref{TQ} reduces to the Bethe ansatz equation
of the integrable higher spin XXZ chain
\begin{align}
\Bigg( \frac{\mathrm{sh}(u_j+s \eta)}{\mathrm{sh}(u_j-s \eta)} \Bigg)^M
=\prod_{\substack{k=1 \\  k \neq j}}^p
\frac{\mathrm{sh}(u_j-u_k+\eta)}{\mathrm{sh}(u_j-u_k-\eta)},
\label{BAE}
\end{align}
which implies that the parameters $\{ u_j \}$
correspond to the Bethe roots. 
Namely, the $TQ$ equation implies the Bethe ansatz equation.

For the half-integer spin $s=L/2 \ (L=1,3,5, \cdots)$ XXZ chain
with odd number of total sites $M=2N+1 \ (N=1,2,3, \cdots)$,
the transfer matrix was found to have simple
eigenvalues at the so-called Razumov-Stroganov point $\eta=-i (L+1) \pi/(L+2)$.
In the sector of $p$ Bethe roots where $NL \le p \le (N+1)L$
(in the sector of total spin $S^z=ML/2-p$),
the transfer matrix was conjectured to have an exact eigenvalue
in the following form \cite{DST}
(obtained by setting $N_k=0,M=0,\ell=1$ in Ref. 42)
\begin{align}
T(u)=2 \mathrm{ch} \Bigg( \frac{(L+1)(ML-2p) \pi i}{2(L+2)} \Bigg) 
\mathrm{sh}^M (u),
\label{transfereigenvalue}
\end{align}
or simply $T(u)=2 \mathrm{ch}(\eta S^z) \mathrm{sh}^M(u)$.
We examine the eigenvalue of the $Q$ operator
corresponding to the eigenvalue \eqref{transfereigenvalue}
of the transfer matrix in two ways.
First, we make an analysis by expanding the $Q$ operator in terms of the
symmetric polynomials of the Bethe roots, regarding them as 
unknown parameters to be solved.
For simplicity, we consider the sector $p=N+M(L-1)/2$
($S^z=1/2$). The other sectors $NL \le p \le (N+1)L$
can be examined in the same way.

We introduce $z=\mathrm{exp}(2u), z_j=\mathrm{exp}(2u_j),
q=\mathrm{exp}(\pi i/(L+2))$ and
redefine the eigenvalue of the $Q$ operator as
\begin{align}
Q(z)=\prod_{j=1}^p (z-z_j),
\label{qoperator}
\end{align}
for convenience.
The $TQ$ equation \eqref{TQ} can be rewritten as
\begin{align}
&-2 \mathrm{ch} \Bigg( \frac{(L+1) \pi i}{2(L+2)} \Bigg)(z-1)^{2N+1}
\prod_{j=1}^p(z-z_j) \nonumber \\
&+
q^{-(L+1)/2}(z-q^2)^{2N+1}
\prod_{j=1}^p(z-q^2 z_j)
+
q^{(L+1)/2}(z-q^{-2})^{2N+1}
\prod_{j=1}^p(z-q^{-2} z_j)=0.
\label{TQone}
\end{align}
Making the expansion of the product of polynomials
in terms of the symmetric polynomials 
\begin{align}
\prod_{j=1}^K(z-z_j)&=\sum_{j=0}^K (-1)^j z^{K-j} e_j, \\
e_j&=\sum_{1 \le i_1 < i_2 < \cdots < i_j \le K}
z_{i_1} z_{i_2} \cdots z_{i_j} \ (j=1,2,\cdots,K), \ e_0=1, \\
(z-1)^K&=\sum_{j=0}^K (-1)^j z^{K-j} \binom{K}{j}, \label{binomialexpansion}
\end{align}
and inserting into \eqref{TQone}, we have
\begin{align}
&2 \sum_{k=0}^{2N+1} \sum_{j=0}^p
(-1)^{j+k} z^{2N+1+p-(j+k)} e_j
\binom{2N+1}{k}
\Bigg\{
-\mathrm{ch} \Bigg( \frac{(L+1)\pi i}{2(L+2)} \Bigg)
+\mathrm{ch} \Bigg( \frac{2 \pi i}{L+2} \Bigg(j+k-\frac{L+1}{4} \Bigg) \Bigg)
\Bigg\}=0.
\end{align}
Changing the summation from $j$ and $k$ to $j$ and $\ell=j+k$, one gets
\begin{align}
&2 \sum_{\ell=0}^{2N+1+p} (-1)^\ell z^{2N+1+p-\ell}
\Bigg\{
-\mathrm{ch} \Bigg( \frac{(L+1)\pi i}{2(L+2)} \Bigg)
+\mathrm{ch} \Bigg( \frac{2 \pi i}{L+2} \Bigg(\ell-\frac{L+1}{4} \Bigg) \Bigg)
\Bigg\}
\sum_{j=\mathrm{max}(0,\ell-2N-1)}^{\mathrm{min}(p,\ell)}
\binom{2N+1}{\ell-j} e_j=0.
\end{align}
Since this equation must hold for arbitrary $z$,
each coefficient of $z^{2N+1+p-\ell}$ has to be zero,
leading to
\begin{align}
&\Bigg\{
-\mathrm{ch} \Bigg( \frac{(L+1)\pi i}{2(L+2)} \Bigg)
+\mathrm{ch} \Bigg( \frac{2 \pi i}{L+2} \Bigg(\ell-\frac{L+1}{4} \Bigg) \Bigg)
\Bigg\}
\sum_{j=\mathrm{max}(0,\ell-2N-1)}^{\mathrm{min}(p,\ell)}
\binom{2N+1}{\ell-j} e_j=0. \label{rewriteTQ}
\end{align}
To obtain the $Q$ operator means to
evaluate the symmetric polynomials of the Bethe roots,
and the problem reduces to solving the linear equations.
Note we have $p$ parameters 
$e_j \ (j=1,2, \cdots,p)$ to be solved.
Taking into account the factor
\begin{align}
-\mathrm{ch} \Bigg( \frac{(L+1)\pi i}{2(L+2)} \Bigg)
+\mathrm{ch} \Bigg( \frac{2 \pi i}{L+2} \Bigg(\ell-\frac{L+1}{4} \Bigg) \Bigg),
\label{factor}
\end{align}
in front of the equation \eqref{rewriteTQ},
we find the equations \eqref{rewriteTQ} with
$\ell=(L+2)k,(L+2)k+(L+1)/2 \ (k=0,1,\cdots,N)$ are automatically satisfied.
Moreover, this factor depends only on $\ell$,
and we find the problem of computing the $Q$ operator eigenvalue
is to solve
the set of $p$ linear equations
\begin{align}
\sum_{j=\mathrm{max}(0,\ell-2N-1)}^{\mathrm{min}(p,\ell)}
\binom{2N+1}{\ell-j} e_j=0, \label{rewriteTQ2}
\end{align}
for 
$\ell=0,1,\cdots,N(L+2)+(L+1)/2 \ 
(\ell \neq (L+2)k,(L+2)k+(L+1)/2 \ (k=0,1,\cdots,N))$.
The number of linear equations is exactly $p$,
the same with that of the paramaters to be solved.
Although the problem of linear dependence remains, many examples
convince us that the equations are linearly independent.
For example, for $L=3$ and $N=3$, the set of linear equations are
\begin{align}
&7+e_1=0,  \nonumber \\
&35+21e_1+7e_2+e_3=0, \nonumber \\
&35+35e_1+21e_2+7e_3+e_4=0, \nonumber \\
&7+21e_1+35e_2+35e_3+21e_4+7e_5+e_6=0, \nonumber \\
&e_1+7e_2+21e_3+35e_4+35e_5+21e_6+7e_7+e_8=0, \nonumber \\
&e_2+7e_3+21e_4+35e_5+35e_6+21e_7+7e_8+e_9=0, \nonumber \\
&e_4+7e_5+21e_6+35e_7+35e_8+21e_9+7e_{10}=0, \nonumber \\
&e_6+7e_7+21e_8+35e_9+35e_{10}=0, \nonumber \\
&e_7+7e_8+21e_9+35e_{10}=0, \nonumber \\
&e_9+7 e_{10}=0. \nonumber
\end{align}
Solving these equations, we easily have
\begin{align}
Q(z)=&\prod_{j=1}^{10}(z-z_j)
=\sum_{j=0}^{10} (-1)^j z^{10-j} e_j \nonumber \\
=&z^{10}+7z^9+\frac{609}{26}z^8
+\frac{1351}{26}z^7+\frac{1064}{13}z^6+\frac{1229}{13}z^5
\nonumber \\
&
+\frac{1064}{13}z^4+\frac{1351}{26}z^3+\frac{609}{26}z^2+7z+1.
\end{align}
We can see that the coefficients of the $Q$ operator are all positive.
Another interesting point we observe is that all the coefficients of the 
linear equations have the form $\binom{2N+1}{j} \ (j=0,1,\cdots,2N+1)$,
which is independent of the spin-$L/2$ but depends on the
total number of sites $M=2N+1$.
The approach made in this section can solve any $Q$ operator in principle
since the problem reduces to solving the linear equations
whose number is the same as that of the parameters to be solved.
This approach might also be useful for
making a guess on the eigenvalue of the transfer matrix.
In general, the number of linear equations
is larger than that of the parameters to be solved.
However, by taking appropriate eigenvalue which can produce
factors such as \eqref{factor}, the number of linear equations
reduces to that of the parameters, which can be solved in a unique way.
So making a guess on the transfer matrix eigenvalue can be
reduced to that on the set of factors like \eqref{factor} which give the 
correct number of zeros.
In the next section, we evaluate the $Q$ operator in a different way.
\section{Spin-3/2}
We can evaluate the $Q$ operator in closed form by
using the interpolation formula
which has been done for the spin-1/2 XXZ chain \cite{FSZ}.
We first consider the spin-3/2 ($L=3$) and $p=3N+1$ Bethe roots ($S^z=1/2$),
since this is the simplest nontrivial and illustrative
example to treat the $Q$ operators
of the higher half-integer spin XXZ chain. 
The other cases can be treated in the same way.
First, let us set 
$f(u)=\mathrm{sh}^{2N+1}(u) \prod_{j=1}^{3N+1} \mathrm{sh}(u-u_j)$.
The $TQ$ equation can be written as
\begin{align}
-2 \mathrm{ch} \Bigg( \frac{2 \pi i}{5} \Bigg) f(u)
+f \Bigg(u+\frac{4}{5} \pi i \Bigg)
+f \Bigg(u+\frac{6}{5} \pi i \Bigg)=0. \label{TQvertwo}
\end{align}
$f(u)$ is a trigonometric polynomial of degree $5N+2$ and satisfies
$f(u+\pi i)=(-1)^N f(u)$.
We make an assumption on $Q(u)=\prod_{j=1}^{3N+1} \mathrm{sh}(u-u_j)$
that it is an even function of $u$
and therefore $f(u)$ is an odd function.
In terms of the symmetric polynomials $e_k$ of $z_j$,
this is equivalent to the assumption $e_{p-j}=(-1)^p e_j$.
The fact that this assumption holds is supported by
solving a set of linear equations \eqref{rewriteTQ2}
in the last section.
Let us make some comment on this assumption.
We can show $e_j=e_j|_{z_k \rightarrow z_k^{-1}}$
which follows from the $z \leftrightarrow z^{-1}, z_j \leftrightarrow z_j^{-1}$
invaiance of the $TQ$ equation. Namely, rewriting \eqref{TQone} as
\begin{align}
&-2 \mathrm{ch} \Bigg( \frac{(L+1) \pi i}{2(L+2)} \Bigg)(z^{-1}-1)^{2N+1}
\prod_{j=1}^p(z^{-1}-z_j^{-1}) \nonumber \\
&+
q^{-(L+1)/2}(z^{-1}-q^2)^{2N+1}
\prod_{j=1}^p(z^{-1}-q^2 z_j^{-1})
+
q^{(L+1)/2}(z^{-1}-q^{-2})^{2N+1}
\prod_{j=1}^p(z^{-1}-q^{-2} z_j^{-1})=0,
\end{align}
we find the symmetric polynomials $e_j|_{z_k \rightarrow z_k^{-1}}$
satisfy exactly the same set of linear equations as $e_j$
by performing the same analysis in the last section.
Using this property, we find the problem of showing
the assumption $e_{p-j}=(-1)^p e_j$
reduces to showing just only one of them $e_p=(-1)^p$
since $e_p^{-1}e_j=e_{p-j}|_{z_k \rightarrow z_k^{-1}}= e_{p-j}$.
\\
From the above properties, we find $f(u)$  has the following form
\begin{align}
f(u)=\sum_{ \substack{ j=-5N/2 \delta_{N,\mathrm{even}} \\
-(5N-1)/2 \delta_{N, \mathrm{odd}}}}^1
a_j \mathrm{sh}(5N+2j)u.
\end{align}
Inserting this expression into \eqref{TQvertwo}, one gets
\begin{align}
2 \sum_{ \substack{ j=-5N/2 \delta_{N,\mathrm{even}} \\
-(5N-1)/2 \delta_{N, \mathrm{odd}}}}
\Bigg\{
-\mathrm{ch} \Bigg( \frac{2 \pi i}{5} \Bigg)
+\mathrm{ch} \Bigg( \frac{2 \pi ji}{5} \Bigg)
\Bigg\}
a_j \mathrm{sh}(5N+2j)u=0.
\end{align}
We easily find
\begin{align}
-\mathrm{ch} \Bigg( \frac{2 \pi i}{5} \Bigg)
+\mathrm{ch} \Bigg( \frac{2 \pi ji}{5} \Bigg)
\neq 0, \ j=0,2,3 \ (\mathrm{mod} \ 5),
\end{align}
which implies $a_j=0$ for $j=0,2,3 \ (\mathrm{mod} \ 5)$,
and the number of  coefficients to be determined reduces.
The function $f(u)$ can be expressed as
\begin{align}
f(u)=\sum_{k=0}^{N/2} \alpha_k \mathrm{sh}(5N+2-10k)u
+\sum_{k=0}^{N/2-1} \beta_k \mathrm{sh}(5N-2-10k)u,
\end{align}
for $N$ even and
\begin{align}
f(u)=\sum_{k=0}^{(N-1)/2} \alpha_k \mathrm{sh}(5N+2-10k)u
+\sum_{k=0}^{(N-1)/2} \beta_k \mathrm{sh}(5N-2-10k)u,
\end{align}
for $N$ odd.
From now on, we only consider the case $N$ even.
The case $N$ odd can be treated in the same way.

The function $f(u)$ is proportional to
$\mathrm{sh}^{M}u$ and therefore
\begin{align}
\Bigg(
\frac{\partial^{\mu}}{\partial u^{\mu}} f(u)
\Bigg)
\Bigg|_{u=0}=0, \ \mu=0,1,\cdots,2N,
\end{align}
follows. Writing down this condition explicitly, we have
\begin{align}
\sum_{k=0}^{N/2} \alpha_k (5N+2-10k)^{2 \mu+1}
+\sum_{k=0}^{N/2-1} \beta_k (5N-2-10k)^{2 \mu+1}=0, \ \mu=0,1,\cdots,N-1.
\label{condition}
\end{align}
This condition
\eqref{condition} is equivalent to the one
that the parameters $\alpha_k \ (k=0,1,\cdots,N/2), \ \beta_k \ 
(k=0,1,\cdots,N/2-1) $ must satisfy the
relation
\begin{align}
\sum_{k=0}^{N/2} \alpha_k (5N+2-10k) W((5N+2-10k)^2)
+\sum_{k=0}^{N/2-1} \beta_k (5N-2-10k) W((5N-2-10k)^2)=0,
\label{conditiontwo}
\end{align}
for any polynomial $W(x)$ of degree equal to or less than $N-1$. \\
We now consider the more general problem \cite{FSZ}:
given a set of different complex numbers $\{x_1,x_2,\dots,x_K \}$,
is there any set of complex numbers $\{\gamma_1,\gamma_2,\dots,\gamma_K \}$
$(\gamma_i \neq 0$ for some $i$)
satisfying the relation
\begin{align}
\sum_{k=1}^K \gamma_k W(x_k)=0,
\end{align}
for any polynomial $W(x)$ of degree equal to or less than $K-2$?
The answer \cite{FSZ} is that up to an overall constant $C$,
there is a unique set of complex numbers
$\gamma_k \ (k=1,2,\cdots,K)$ which can be expressed as
\begin{align}
\gamma_k=\frac{C}{\prod_{\substack{j=1 \\ j \neq k}}^K(x_k-x_j)}.
\end{align}
This formula comes from the equality
\begin{align}
\sum_{k=1}^K \frac{x_k^\ell}{\prod_{\substack{j=1 \\ j \neq k}}^K(x_k-x_j)}
=0, \ 0 \le \ell \le K-2, \label{interaux}
\end{align}
which can be shown as follows \cite{Toda,JJ}.
Consider a polynomial $V(x)=\prod_{j=1}^K(x-x_j)$ and the following integral
\begin{align}
\frac{1}{2 \pi i} \oint \frac{z^{\ell+1} d z}{V(z)(z-x)}, \ 0 \le \ell \le K-2,
\end{align}
where the contour surrounds $x$ and all the zeros of $V(x)$.
Taking the residues, one gets
\begin{align}
\frac{1}{2 \pi i} \oint \frac{z^{\ell+1} d z}{V(z)(z-x)}
=\frac{x^{\ell+1}}{V(x)}
+\sum_{k=1}^K \frac{x_k^{\ell+1}}{V^\prime(x_k)(x_k-x)}.
\end{align}
On the other hand, regarding the contour as a circle at infinity,
the integral becomes zero for $0 \le \ell \le K-2$,
and we have
\begin{align}
\frac{x^{\ell+1}}{V(x)}
+\sum_{k=1}^K \frac{x_k^{\ell+1}}{V^\prime(x_k)(x_k-x)}=0.
\end{align}
Taking $x=0$, one has \eqref{interaux}.

Applying the above fact to the
special case \eqref{conditiontwo} which we consider, one gets
\begin{align}
&\alpha_k(5N+2-10k) \nonumber \\
=&\frac{C}
{
\prod_{\substack{j=0 \\ j \neq k}}^{N/2}
\{(5N+2-10k)^2-(5N+2-10j)^2 \}
\prod_{j=0}^{N/2-1}
\{(5N+2-10k)^2-(5N-2-10j)^2 \}
}, \nonumber \\
&\beta_k(5N-2-10k) \nonumber \\
=&\frac{C}
{
\prod_{j=0}^{N/2}
\{(5N-2-10k)^2-(5N+2-10j)^2 \}
\prod_{\substack{j=0 \\ j \neq k}}^{N/2-1}
\{(5N-2-10k)^2-(5N-2-10j)^2 \}
},
\end{align}
with some factor $C$, which can be determined by the coefficient
of $\mathrm{exp}(5N+2)u$ of $f(u)$
\begin{align}
\alpha_0=\frac{\prod_{j=1}^{3N+1} \mathrm{e}^{-u_j}}{2^{5N+1}}.
\end{align}
After various simplifications,
one gets
\begin{align}
\Bigg( \prod_{j=1}^{3N+1} \mathrm{e}^{u_j} \Bigg) \alpha_k
&=\frac{(-1)^k }{2^{5N+1}} \binom{N}{k} \prod_{j=0}^N \frac{2+5j}{2-5k+5j}, \\
\Bigg( \prod_{j=1}^{3N+1} \mathrm{e}^{u_j} \Bigg) \beta_k
&
=\frac{(-1)^k }{2^{5N+1}} \binom{N}{k} \prod_{j=0}^N \frac{2+5j}{-2-5k+5j},
\end{align}
and the function $f(u)$ has the following form
\begin{align}
\Bigg( \prod_{j=1}^{3N+1} \mathrm{e}^{u_j} \Bigg) f(u)=&
\sum_{k=0}^{N/2}
\Bigg(
\frac{(-1)^k }{2^{5N+1}} \binom{N}{k} \prod_{j=0}^N \frac{2+5j}{2-5k+5j}
\Bigg)
\mathrm{sh}(5N+2-10k)u \nonumber \\
&+\sum_{k=0}^{N/2-1}
\Bigg(
\frac{(-1)^k }{2^{5N+1}} \binom{N}{k} \prod_{j=0}^N \frac{2+5j}{-2-5k+5j}
\Bigg)
\mathrm{sh}(5N-2-10k)u. \label{preqoperator}
\end{align}
The expression \eqref{preqoperator} leads to the following
expression for the $Q$ operator \eqref{qoperator} for $N$ even
\begin{align}
Q(z)=&(z-1)^{-(2N+1)}
\Bigg\{
\sum_{k=0}^{N/2}
\Bigg(
(-1)^k \binom{N}{k} \prod_{j=0}^N \frac{2+5j}{2-5k+5j}
\Bigg)
(z^{5N+2-5k}-z^{5k}) \nonumber \\
&+
\sum_{k=0}^{N/2-1}
\Bigg(
(-1)^k \binom{N}{k} \prod_{j=0}^N \frac{2+5j}{-2-5k+5j}
\Bigg)
(z^{5N-5k}-z^{5k+2})
\Bigg\}.
\end{align}
The $Q$ operator for $N$ odd is
\begin{align}
Q(z)=&(z-1)^{-(2N+1)}
\Bigg\{
\sum_{k=0}^{(N-1)/2}
\Bigg(
(-1)^k \binom{N}{k} \prod_{j=0}^N \frac{2+5j}{2-5k+5j}
\Bigg)
(z^{5N+2-5k}-z^{5k}) \nonumber \\
&+
\sum_{k=0}^{(N-1)/2}
\Bigg(
(-1)^k \binom{N}{k} \prod_{j=0}^N \frac{2+5j}{-2-5k+5j}
\Bigg)
(z^{5N-5k}-z^{5k+2})
\Bigg\}.
\end{align}
One can read off the $Q$ operator at special points of $z$
from this expression.
For example,
$Q(0)=1$ for both $N$ even and odd, and
\begin{align}
Q(-1)=&\left\{
\begin{array}{cc}
0, & N \ \mathrm{even}, \\
4^{-N} \sum_{k=0}^{(N-1)/2} \binom{N}{k}
\Bigg(
\prod_{j=0}^N \frac{2+5j}{2-5k+5j}+\prod_{j=0}^N \frac{2+5j}{-2-5k+5j}
\Bigg),
& N \ \mathrm{odd}.
\end{array}
\right.
\end{align}
\section{Spin-$L/2$}
The steps to calculate the $Q$ operator in the last section can be applied
to the general case of the higher half-integer spin XXZ chain as well.
Recall that we are considering
the half-integer spin-$s=L/2 \ (L=1,3,5, \cdots)$ XXZ chain
with odd sites $M=2N+1 \ (N=1,2,3, \cdots)$ at the
Razumov-Stroganov point $\eta=-i (L+1) \pi/(L+2)$
on the sector of $p$ Bethe roots where $NL \le p \le (N+1)L$.
Repeating the same steps as the last section, we find
\begin{align}
Q(z)=&(z-1)^{-(2N+1)}
\Bigg\{
\sum_{k=0}^{N/2}
\Bigg(
(-1)^k \binom{N}{k} \prod_{j=0}^N 
\frac{p+1-LN+(L+2)j}{p+1-LN+(L+2)(j-k)}
\Bigg)
(z^{p+1+2N-(L+2)k}-z^{(L+2)k}) \nonumber \\
&+
\sum_{k=0}^{N/2-1}
\Bigg(
(-1)^k \binom{N}{k} \prod_{j=0}^N 
\frac{p+1-LN+(L+2)j}{LN-p-1+(L+2)(j-k)}
\Bigg)
(z^{(L+2)(N-k)}-z^{p+1-LN+(L+2)k})
\Bigg\},
\end{align}
for $N$ even and
\begin{align}
Q(z)=&(z-1)^{-(2N+1)}
\Bigg\{
\sum_{k=0}^{(N-1)/2}
\Bigg(
(-1)^k \binom{N}{k} \prod_{j=0}^N 
\frac{p+1-LN+(L+2)j}{p+1-LN+(L+2)(j-k)}
\Bigg)
(z^{p+1+2N-(L+2)k}-z^{(L+2)k}) \nonumber \\
&+
\sum_{k=0}^{(N-1)/2}
\Bigg(
(-1)^k \binom{N}{k} \prod_{j=0}^N 
\frac{p+1-LN+(L+2)j}{LN-p-1+(L+2)(j-k)}
\Bigg)
(z^{(L+2)(N-k)}-z^{p+1-LN+(L+2)k})
\Bigg\},
\end{align}
for $N$ odd.
From these expressions, we can see $Q(0)=1$ for both $N$ even and odd.
One can also find that the $Q(z)$ has a zero at $z=-1$
for $L$ and $p$ odd.

\section{Functional relations}
One can derive the hierarchy of functional relations by combining the
$Q$ operator on the ``right side" 
and its corresponding ``wrong side" 
as the spin-1/2 XXZ chain and the conformal field theory
\cite{KLWZ,BLZ2,PS} by using Pl\"ucker relations.
The ``right side" (``wrong side") means the sector with positive 
(negative) total spin.
From now on, we denote the $Q$ operator on the ``right side"
by $Q(z)$, and
the $Q$ operator on the corresponding ``wrong sector"
by $P(z)$.
The two sectors have the same exact transfer matrix eigenvalue.
We rewrite the two $TQ$ equations in the following form
\begin{align}
2 \mathrm{ch} \Bigg( \frac{(L+1)(ML-2p) \pi i}{2(L+2)} \Bigg)(z-1)^{2N+1}Q(z)
=&q^{(L+1)(ML-2p)/2}(z q^{-2}-1)^{2N+1}
Q(z q^{-2}) \nonumber \\
&+q^{-(L+1)(ML-2p)/2}(z q^2-1)^{2N+1}
Q(z q^2), \label{TQfun} \\
2 \mathrm{ch} \Bigg( \frac{(L+1)(ML-2p) \pi i}{2(L+2)} \Bigg)(z-1)^{2N+1}P(z)
=&q^{-(L+1)(ML-2p)/2}(z q^{-2}-1)^{2N+1}
P(z q^{-2}) \nonumber \\
&+q^{(L+1)(ML-2p)/2}(z q^2-1)^{2N+1}
P(z q^2). \label{TQfuntwo}
\end{align}
Eliminating the transfer matrix eigenvalue
$2 \mathrm{ch} \Bigg( \frac{(L+1)(ML-2p) \pi i}{2L} \Bigg)(z-1)^{2N+1}$
by combining the two equations \eqref{TQfun} and \eqref{TQfuntwo}
leads to
\begin{align}
&(z q^{-2}-1)^{2N+1}
\{
q^{(L+1)(ML-2p)/2} Q(z q^{-2})P(z)
-q^{-(L+1)(ML-2p)/2} P(z q^{-2})Q(z)
\} \nonumber \\
=&
(z q^2-1)^{2N+1}
\{
q^{(L+1)(ML-2p)/2} P(z q^2)Q(z)
-q^{-(L+1)(ML-2p)/2} Q(z q^2)P(z)
\},
\end{align}
therefore we can define a polynomial $\Psi_1(z)$ of degree 
$(2N+1)(L-1)$ satisfying
\begin{align}
q^{(L+1)(ML-2p)/2} P(z q^2)Q(z)
-q^{-(L+1)(ML-2p)/2} Q(z q^{2})P(z)
&=(z q^{-2}-1)^{2N+1} \Psi_1(z), \label{PQone}  \\
q^{(L+1)(ML-2p)/2} Q(z q^{-2})P(z)
-q^{-(L+1)(ML-2p)/2} P(z q^{-2})Q(z)
&=(z q^2-1)^{2N+1} \Psi_1(z) \label{PQtwo}.
\end{align}
Shifting $z \to z q^{-1} $ in \eqref{PQone}
and $z \to z q$ in \eqref{PQtwo} leads to
\begin{align}
&q^{(L+1)(ML-2p)/2} 
P(z q)Q(z q^{-1})
-q^{-(L+1)(ML-2p)/2} 
Q(z q)P(z q^{-1})
\nonumber \\
=&(z q^{-3}-1)^{2N+1} \Psi_1(z q^{-1})
\nonumber \\
=&(z q^{3}-1)^{2N+1} \Psi_1(z q).
\label{PQthree}
\end{align}
The equality \eqref{PQthree} leads to the
existence of the polynomial $\Psi_2(z)$ of degree $(2N+1)(L-2)$
satisfying
\begin{align}
\Psi_1(z q^{-1})&=(z q^3-1)^{2N+1}
\Psi_2(z), \nonumber \\
\Psi_1(z q)&=(z q^{-3}-1)^{2N+1}
\Psi_2(z).
\end{align}
Inserting these expressions into \eqref{PQthree}, we get
\begin{align}
&q^{(L+1)(ML-2p)/2} 
P(z q)Q(z q^{-1})
-q^{-(L+1)(ML-2p)/2} 
Q(z q)P(z q^{-1})
\nonumber \\
=&(z q^{-3}-1)^{2N+1}
(z q^3-1)^{2N+1} \Psi_2(z) \nonumber \\
=&(z q^{-3}-1)^{2N+1}
(z q^{-5}-1)^{2N+1} \Psi_2(z q^{-2})
\nonumber \\
=&(z q^3-1)^{2N+1}
(z q^5-1)^{2N+1} \Psi_2(z q^2).
\label{PQfour}
\end{align}
For the case of spin-3/2 $(L=3)$, \eqref{PQfour} leads to the
determination of $\Psi_2(z)$ as 
$\Psi_2(z)=C (z q^5-1)^{2N+1}$ which finally results in the
following equality
\begin{align}
&q^{2(ML-2p)} 
P(z q)Q(z q^{-1})
-q^{-2(ML-2p)} 
Q(z q)P(z q^{-1})
=C(z q^3-1)^{2N+1}
(z q^5-1)^{2N+1}
(z q^7-1)^{2N+1}.
\end{align}
For generic $L$,
one can repeat the same argument to find
\begin{align}
&q^{(L+1)(ML-2p)/2} 
P(z q)Q(z q^{-1})
-q^{-(L+1)(ML-2p)/2} 
Q(z q)P(z q^{-1})
=C \prod_{j=1}^{L}(z q^{2j+1}-1)^{2N+1}.
\end{align}
Evaluating the both hand sides of this equality at $z=0$ leads to
the determination of the constant factor $C$ as
\begin{align}
C=2 \mathrm{sh} \Bigg( \frac{(L+1)(2p-ML) \pi i}{2(L+2)} \Bigg).
\end{align}
Thus we have
\begin{align}
&q^{(L+1)(ML-2p)} 
P(z q)Q(z q^{-1})
-q^{-(L+1)(ML-2p)/2} 
Q(z q)P(z q^{-1})
=2 \mathrm{sh} \Bigg( \frac{(L+1)(2p-ML) \pi i}{2(L+2)} \Bigg) 
\prod_{j=1}^{L}(z q^{2j+1}-1)^{2N+1},
\label{PQfive}
\end{align}
which is the first fundamental relation between the
$Q$ operator and the $P$ operator.

We next proceed to find the second fundamental relation
which, together with the first one, leads to the construction
of the hierarchy of the functional relations.
Comparing
\eqref{PQthree} and \eqref{PQfive}, we have
\begin{align}
\Psi_1(z)=2 \mathrm{sh} \Bigg( \frac{(L+1)(2p-ML)\pi i}{2(L+2)} \Bigg)
\prod_{j=2}^{L}(z q^{2j}-1)^{2N+1}. \label{psione}
\end{align}
Multiplying the both hand sides of the second $TQ$ equation
\eqref{TQfuntwo} by $\Psi_1(z)$ and using the relations
\eqref{PQone} and \eqref{PQtwo}, we find the following relation
\begin{align}
&2 \mathrm{ch} \Bigg( \frac{(L+1)(2p-ML) \pi i}{2(L+2)} \Bigg)(z-1)^{2N+1}
\Psi_1(z)
=q^{(L+1)(ML-2p)}
P(z q^2) Q(z q^{-2})
-q^{-(L+1)(ML-2p)}
Q(z q^2) P(z q^{-2}).
\end{align}
Inserting \eqref{psione}, one gets the second fundamental
functional relation
\begin{align}
&q^{(L+1)(ML-2p)} 
P(z q^2)Q(z q^{-2})
-q^{-(L+1)(ML-2p)} 
Q(z q^2)P(z q^{-2})
\nonumber \\
=&2 \mathrm{sh} \Bigg( \frac{(L+1)(2p-ML) \pi i}{L+2} \Bigg) 
(z-1)^{2N+1}
\prod_{j=2}^{L}(z q^{2j}-1)^{2N+1}.
\label{PQsix}
\end{align}
The left hand sides of the functional relations
\eqref{PQfive} and \eqref{PQsix} can be regarded as the discrete analogue
of Wronskian.
To obtain the hierarchy of functional relations,
we slightly change from
$Q(z)$ and $P(z)$ to $\tilde{Q}(z)=z^{(L+1)(p/2-(ML-1)/4)}Q(z)$
and $\tilde{P}(z)=z^{(L+1)((ML+1)/4-p/2)}P(z)$
to rewrite the relations \eqref{PQfive} and \eqref{PQsix}
in a more symmetric form
\begin{align}
\tilde{P}(z q)\tilde{Q}(z q^{-1})
-\tilde{Q}(z q)\tilde{P}(z q^{-1})
=&2 \mathrm{sh} \Bigg( \frac{(L+1)(2p-ML) \pi i}{2(L+2)} \Bigg) 
z^{(L+1)/2} \prod_{j=1}^{L}(z q^{2j+1}-1)^{2N+1},
\label{fundone}
\\
\tilde{P}(z q^2)\tilde{Q}(z q^{-2})
-\tilde{Q}(z q^2)\tilde{P}(z q^{-2})
=&2 \mathrm{sh} \Bigg( \frac{(L+1)(2p-ML) \pi i}{L+2} \Bigg) 
z^{(L+1)/2} (z-1)^{2N+1}
\prod_{j=2}^{L}(z q^{2j}-1)^{2N+1}.
\label{fundtwo}
\end{align}
We now define the following family of functions
\begin{align}
t_s(z)=\tilde{P}(z q^{2s+1}) \tilde{Q}(z q^{-(2s+1)})
-\tilde{P}(z q^{-(2s+1)}) \tilde{Q}(z q^{2s+1}),
\end{align}
$(s=0,\pm 1/2, \pm 1, \cdots)$
which is a family of quantum Wronskian.
From the definition and \eqref{fundone}, \eqref{fundtwo},
one finds the function
$t_s(z)$ satisfies
\begin{align}
t_{-s-1}(z)&=-t_s(z), \ t_{-1/2}(z)=0, \\
t_0(z)&=2 \mathrm{sh} \Bigg( \frac{(L+1)(2p-ML) \pi i}{2(L+2)} \Bigg) 
z^{(L+1)/2} \prod_{j=1}^{L}(z q^{2j+1}-1)^{2N+1}, \\
t_{1/2}(z)&=2 \mathrm{sh} \Bigg( \frac{(L+1)(2p-ML) \pi i}{L+2} \Bigg) 
z^{(L+1)/2} (z-1)^{2N+1}
\prod_{j=2}^{L}(z q^{2j}-1)^{2N+1}.
\end{align}
We define the function $\Delta(a,b)$ as
\begin{align}
\Delta(a,b)=\tilde{P}(a)\tilde{Q}(b)-\tilde{Q}(a)\tilde{P}(b),
\end{align}
which satisfies the Pl\"ucker relation
\begin{align}
\Delta(a,b)\Delta(c,d)-\Delta(a,c)\Delta(b,d)+\Delta(a,d)\Delta(b,c)=0.
\label{Pl}
\end{align}
Noting the function $t_s(z)$ can be expressed as
$t_s(z)=\Delta(z q^{2s+1},
z q^{-(2s+1)})$,
one can see that
setting $a=z, \ b=z q^{-2(2s_1+1)}, \ 
c=z q^{-2(2s_2+1)}, \ d=z q^{-2(2s_3+1)}$
in the relation \eqref{Pl} leads to the hierarchy of functional relations
\begin{align}
&t_{s_1}(z q^{-(2s_1+1)})
t_{s_3-s_2-1/2}(z q^{-2(s_2+s_3+1)})
-
t_{s_2}(z q^{-(2s_2+1)})
t_{s_3-s_1-1/2}(z q^{-2(s_1+s_3+1)})
\nonumber \\
+&
t_{s_3}(z q^{-(2s_3+1)})
t_{s_2-s_1-1/2}(z q^{-2(s_1+s_2+1)})=0.
\end{align}
In particular, we have the following fundamental fusion relations
\begin{align}
&2 \mathrm{ch} \Bigg( \frac{(L+1)(2p-ML) \pi i}{2(L+2)} \Bigg)
(z-1)^{2N+1} t_s(z q^{-(2s+1)}) \nonumber \\
=&q^{(L+1)/2} \prod_{j=1}^{L}
(z q^{-2}-1)^{2N+1}
t_{s-1/2}(z q^{-2(s+1)})
+q^{-(L+1)/2} \prod_{j=1}^{L}
(z q^2-1)^{2N+1} t_{s+1/2}(z q^{-2 s}),
\end{align}
from $s_1=s, \ s_2=-1, \ s_3=0$.

\section{Summary and Discussion}
In this paper, we investigated the $TQ$ equation and the $Q$ operator
of higher half-integer integer spin XXZ chain at the Razumov-Stroganov point.
First, by making expansion of the $Q$ operator in terms of the symmetric
polynomials of the Bethe roots and regarding them as unknown parameters,
we showed that solving the $TQ$ equations reduces to solving a simple
set of linear equations.
This approach might be useful for making a guess on the
Razumov-Stroganov point of other models, boundary conditions and so on.
Next, rewriting the conditions that the $Q$ operator must
satisfy to a form such that the interpolation formula can be applied,
the $Q$ operator is evaluated in closed form.
By combining the $Q$ operators on the ``right side" and the ``wrong side",
we showed they produce the hierarchy of functional relations.

The analysis made in this paper is for the models with
the periodic boundary condition.
It should be straightforward to extend the analysis
to open boundary conditions.
It is interesting to lift the analysis to the elliptic case.
For the spin-1/2 case, the groundstate transfer matrix eigenvalue
of the XYZ chain is found to be simply expressed as a theta function 
\cite{Baxter2},
which reduces to a trigonometric function at the trigonometric limit.
This fact and the one for the higher half-integer XXZ chain \cite{DST}
would be reasons to believe that the
exact transfer matrix eigenvalue can be lifted to
the elliptic higher half-integer XYZ spin.
Making a guess on the transfer matrix eigenvalue
and investigating the $TQ$ equation and the $Q$ operators is
an interesting problem to study.

\section{Acknowledgements}
The author thanks the anonymous referee for careful reading and
valuable suggestions for the improvement of the paper.
This work was partially supported by
Grants-in-Aid for Young Scientists (B) No. 25800223
and for Scientific Research (C) No. 24540393.

\appendix

\renewcommand{\theequation}{A\arabic{equation}}
\setcounter{equation}{0}

\section{Relation between the $Q$ operators}

In this appendix, we show the general construction of the
relation between the $Q$ operator $Q(z)$ on the ``right side"
and its corresponding $Q$ operator $P(z)$ on the ``wrong side".
First, we divide the both hand sides of the
first fundamental functional relation \eqref{PQfive} by 
$Q(z q)Q(z q^{-1})$ as
\begin{align}
&q^{(L+1)(ML-2p)/2} 
\frac{P(z q)}{Q(z q)}
-q^{-(L+1)(ML-2p)/2} 
\frac{P(z q^{-1})}{Q(z q^{-1})}
=\frac{2 \mathrm{sh} \Bigg( \frac{(L+1)(2p-ML) \pi i}{2(L+2)} \Bigg) 
\prod_{j=1}^{L}(z q^{2j+1}-1)^{2N+1}}{Q(z q)Q(z q^{-1})}.
\nonumber \\
\label{PQdef}
\end{align}
By partial fraction decomposition, the right hand side of the \eqref{PQdef}
can be expressed as
\begin{align}
\frac{2 \mathrm{sh} \Bigg( \frac{(L+1)(2p-ML) \pi i}{2(L+2)} \Bigg) \prod_{j=1}^{L}(z q^{2j+1}-1)^{2N+1}}{Q(z q)Q(z q^{-1})}
=R(z)+\frac{A(z q)}{Q(z q)}
-\frac{B(z q^{-1})}{Q(z q^{-1})},
\label{decomposition}
\end{align}
with a polynomial $R(z)$ of degree $ML-2p$ and
polynomials $A(z)$ and $B(z)$ of degree less than 
$\mathrm{deg}Q(z)=p$.
Next, dividing the both hand sides of the
second fundamental functional relation \eqref{PQsix} by
$Q(z q^2)Q(z q^{-2})$
and using the decomposition \eqref{decomposition},
one gets
\begin{align}
&\frac{2 \mathrm{sh} \Bigg( \frac{(L+1)(2p-ML) \pi i}{(L+2)} \Bigg) 
(z-1)^{2N+1}
\prod_{j=2}^{L}(z q^{2j}-1)^{2N+1}}
{Q(z q^2)Q(z q^{-2})} \nonumber \\
=&q^{(L+1)(ML-2p)} 
\frac{P(z q^2)}{Q(z q^2)}
-q^{-(L+1)(ML-2p)}
\frac{P(z q^{-2})}{Q(z q^{-2})}
\nonumber \\
=&q^{(L+1)(ML-2p)/2}
R(z q)
+q^{-(L+1)(ML-2p)/2}
R(z q^{-1})
+q^{(L+1)(ML-2p)/2} 
\frac{A(z q^2)}{Q(z q^2)}
-q^{-(L+1)(ML-2p)/2} 
\frac{B(z q^{-2})}{Q(z q^{-2})}
\nonumber \\
&+\frac{
q^{-(L+1)(ML-2p)/2}A(z)
-q^{(L+1)(ML-2p)/2}B(z)
}{Q(z)}. \label{decompositiontwo}
\end{align}
If the last term of the right hand side of \eqref{decompositiontwo}
is not zero, the zeros of $Q(z)$ would be poles of this equality,
which does not exist in the left hand side, leading to contradiction.
Thus the polynomials $A(z)$ and $B(z)$ are related as
\begin{align}
A(z)&=q^{(L+1)(ML-2p)/2}C(z), \nonumber \\
B(z)&=q^{-(L+1)(ML-2p)/2}C(z),
\end{align}
with some polynomial $C(z)$ essentially the same with $A(z)$ and $B(z)$.
We also assume that the polynomial $R(z)$ can be decomposed as
\begin{align}
R(z)=q^{(L+1)(ML-2p)/2} F(z q)
-q^{-(L+1)(ML-2p)/2} F(z q^{-1}),
\end{align}
with some polynomial $F(z)$ whose degree is the same with $R(z)$.
The combination of \eqref{PQdef} and \eqref{decomposition}
now becomes
\begin{align}
&q^{(L+1)(ML-2p)/2} 
\frac{P(z q)}{Q(z q)}
-q^{-(L+1)(ML-2p)/2} 
\frac{P(z q^{-1})}{Q(z q^{-1})}
\nonumber \\
=&q^{(L+1)(ML-2p)/2}
\Bigg(
F(z q)
+\frac{C(z q)}{Q(z q)}
\Bigg)
-q^{-(L+1)(ML-2p)/2} 
\Bigg(
F(z q^{-1})
+\frac{C(z q^{-1})}{Q(z q^{-1})}
\Bigg),
\end{align}
from which one finds the $Q$ operator on the ``wrong side" $P(z)$
should be expressed by the $Q$ operator on the ``right side" $Q(z)$
as
\begin{align}
P(z)&=F(z)Q(z)+C(z) \nonumber \\
&=F(z)Q(z)+q^{-(L+1)(ML-2p)/2}A(z) \nonumber \\
&=F(z)Q(z)+q^{(L+1)(ML-2p)/2}B(z),
\end{align}
where $F(z)$ and $A(z)$ are obtained by the partial fraction
decomposition \eqref{decomposition} of the $Q$ operator $Q(z)$.

\end{document}